\documentstyle[aps,prd]{revtex}
\begin{document}
\tightenlines
\def\be{\begin{equation}}
\def\ee{\end{equation}}
\def\bea{\begin{eqnarray}}
\def\eea{\end{eqnarray}}
\def\nn{\nonumber}
\def\th{\theta}
\def\ph{\phi}
\def\lt{\left}
\def\rt{\right}
\def\degree{\mathop{\rm {{}^\circ}}}
\input epsf.tex
\title{Schwarzschild black hole lensing}
\author{K. S. Virbhadra\thanks{Email address : shwetket@maths.uct.ac.za}
     and George F. R. Ellis\thanks{Email address  : ellis@maths.uct.ac.za}
}
\address{Department of Applied Mathematics,
                    University of Cape Town,
                    Rondebosch 7701,
                    South Africa                
                    }
\maketitle
\begin{abstract} 
   We study strong gravitational lensing due to a  Schwarzschild black hole. 
Apart from the primary and the secondary images we find a sequence of images 
on both sides of the optic axis; we call them {\em relativistic images}. These
images are formed due to large bending of light  near  $r = 3M$ (the closest 
distance of approach $r_o$ is greater than $3M$). The sources  of the entire 
universe are mapped in the vicinity of the  black hole by these images. For the
case of the Galactic supermassive ``black  hole'' they  are formed at about 
$17$ microarcseconds from the optic axis. The relativistic images are  not 
resolved among  themselves, but they are resolved from the primary and 
secondary images.  However the  relativistic images are very much demagnified  
unless the observer, lens and  source  are very highly aligned. Due to this
and some  other  difficulties the observation of these images does not
seem to be feasible in near future. However, it  would be  a great success  of
the general theory of relativity in a   strong  gravitational field if they
ever were observed and it would also give an  upper bound, $r_o = 3.21 M$, to 
the compactness of the lens, which would support the black hole interpretation 
of the lensing  object. 
\end{abstract}

\pacs{04.70Bw, 97.60Lf, 98.62Sb, 04.80Cc}

\section{Introduction}
\label{sec:intro}
The phenomena resulting from  the deflection of electromagnetic radiation in
a gravitational field are referred to as  {\em gravitational lensing} (GL)
and an object causing a detectable deflection is  known as a {\em gravitational 
lens}.  The basic theory of GL was developed by Liebes\cite{Lie64}, 
Refsdal\cite{Ref64}, and Bourossa and Kantowski\cite{BK75}.
For detailed discussions on GL see the  monograph by Schneider {\it et al.}
\cite{Schetal92} and  reviews by  Blandford and Narayan\cite{BN92}, 
Refsdal and Surdej\cite{RS94}, Narayan and Bartelmann\cite{NB96} and 
Wambsganss\cite{Wam98}.

The discovery of quasars in 1963 paved the way for observing  point
source GL.  Walsh, Carswell and Weymann\cite{WCW79} discovered the first 
example of  GL. They observed twin images  QSO 0957+561 A,B separated by $5.7$ 
arcseconds at the same redshift $z_s = 1.405$ and mag $\approx 17$. 
Following this remarkable discovery  more than  a dozen convincing multiple-imaged 
quasars are known. 

The  vision of Zwicky that galaxies can be lensed was crystallized when
Lynds and Petrosian\cite{LP86} and Soucail {\it et al.}\cite{Souetal87}
independently observed
giant blue luminous {\em arcs} of about $20$ arcseconds long in the rich 
clusters of galaxies. Paczy\'{n}ski\cite{Pac87}  interpreted these  
giant arcs to be distorted images of distant galaxies located behind the
clusters. About $20$ giant arcs  have been observed in the rich clusters.  
Apart from the giant arcs, there have been also observed weakly distorted 
{\em arclets} which are images of other faint background 
galaxies\cite{Tys88}.

Hewitt {\it et al.}\cite{Hew88}  observed the first Einstein ring MG1131+0456 at 
redshift $z_s = 1.13$. With high resolution radio observations, they 
found the extended radio source to  actually be  a ring of diameter 
about $1.75$ {\em arcseconds}.   There are about half a dozen observed rings
of diameters between $0.33$ to $2$ arcseconds and all of them are found in 
the radio waveband;  some  have optical and infrared counterparts as 
well\cite{Wam98}. 

The general theory of relativity has passed  experimental tests
in a weak gravitational field with flying colors; however, the theory
has not been tested  in a strong gravitational field.  Testing the 
gravitational field in the vicinity of a compact massive object, such as 
a black hole or a neutron star, could be a possible avenue for such
investigations. Dynamical 
observations of several galaxies  show that their centres contain 
massive dark objects. Though there is no iron-clad evidence, indirect 
arguments suggest that these are supermassive black holes; at least, the 
case for black holes in the Galaxy as well as in NGC4258  appears to be strong 
\cite{Ricetal98}. These could be possible observational targets to test 
the Einstein theory of relativity in a strong gravitational field through GL.

Immediately after the advent of the  general theory of relativity, 
Schwarzschild obtained a static spherically symmetric asymptotically 
flat vacuum solution to  the Einstein  equations, which was later found to 
have an event horizon when maximally extended; thus this solution represents 
the  gravitational field of a spherically symmetric black hole (see in Hawking
and Ellis\cite{HE73}).
Schwarzschild  GL in the weak gravitational field region (for which the 
deflection  angle  is  small)  is well-known\cite{Schetal92}. 
Recently Kling {\it et al.}\cite{KNP99} developed an iterative approach to GL
theory based on approximate solutions of the null geodesics equations,
and to illustrate their method they constructed the iterative lens
equations and time of arrival equation for a single Schwarzschild lens.
In this paper we obtain  a lens equation that allows for the   large
bending of light near a black hole, model the Galactic  supermassive 
``black hole'' as a Schwarzschild lens and  study  point source  lensing 
in  the strong  gravitational field region, when the bending angle can be 
very large.  Apart from a  primary image  and a secondary image 
(which are observed
due to small bending of light in a weak gravitational field)
we get a theoretically infinite sequence of images on both sides  
close to  the optic axis; we term them {\em 
relativistic  images}.  The relativistic images are formed due to large
bending of light  in a strong gravitational field in the vicinity of $3M$, 
and are usually greatly demagnified (the magnification decreases very fast 
with an increase in 
the angular position of the source from the optic axis).  Though the 
observation of relativistic images is a very
difficult  task (it is very unlikely that they will be observed in near 
future), if it ever were  accomplished  it would support the general 
theory of  relativity in a strong gravitational field inaccessible to test 
the theory in any other known way and would also give an upper bound to the
compactness of the lens. This is the subject of study in this  paper. We use 
geometrized units (the  gravitational constant $G = 1$ and  the speed of light 
in vacuum $c = 1$ so that $ M \equiv M G / c^2$).
\section{ Lens equation, magnification and critical curves}
\label{sec:lenseqn}
In this section we derive a lens equation that allows for the   large
bending of light near a black hole.
The lens diagram is given in Fig.1. The line joining the observer $O$ and
the lens $L$ is taken as the reference (optic) axis. The spacetime  under
consideration, with the lens (deflector) causing strong curvature,  is 
asymptotically flat; the observer as well as the  source are situated in the 
flat spacetime region (which can be embedded in an expanding Robertson-Walker
universe).
   \begin{figure*}
\epsfxsize 6cm
   \epsfbox{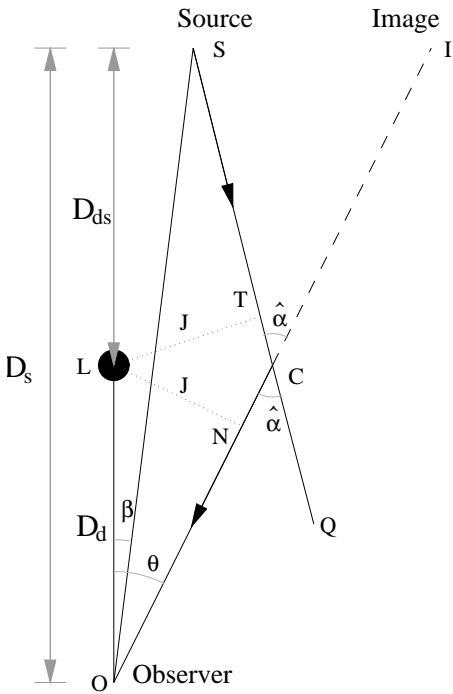}

 \caption[ ]
     { The  lens diagram: 
$O, L$ and $S$ are respectively the positions of the observer, deflector 
(lens) and source. $OL$ is the reference (optic) axis. $\angle LOS$ and 
$\angle LOI$ are the angular separations  of the source and the image 
from the optic axis. $SQ$  and $OI$ are respectively   tangents to the 
null geodesic
at the source and  observer positions; $LN$ and $LT$, the perpendiculars 
to  these tangents  from $L$, are the impact parameter $J$. $\angle OCQ$, 
is the Einstein bending angle. $D_{s}$ 
represents the observer-source  distance, $D_{ds}$ the lens-source distance
and $D_{d}$ the observer-lens distance.
          }
\label{fig1}
\end{figure*}

$SQ$ and $OI$ are tangents to the null geodesic at the source and image 
positions, respectively;  $C$ is where their point of intersection would be
if there were no lensing object present. The angular 
positions  of the source and the image are measured from the optic axis $OL$. 
$\angle LOI$ (denoted by $\theta$)  is the image position 
and $\angle LOS$ (denoted by $\beta$) is the source position if there were no
lensing object. $\hat{\alpha}$ (i.e.  $\angle OCQ$) is the Einstein deflection 
angle. 
The null geodesic  and the background broken geodesic path $OCS$ will be almost
identical, except near the lens where most of bending will take place. Given 
the  vast distances from  observer to lens and from lens to  source, 
this  will be a good approximation, even if the light goes round and round the
lens before reaching the observer.
We assume that the line joining the point $C$  and the location of the lens $L$
is perpendicular to the optic axis. This is a good approximation for small 
values  of $\beta$. We draw perpendiculars $LT$ and $LN$ from $L$ on the 
tangents $SQ$ and $OI$ repectively and these represent the impact parameter 
$J $. $D_s$ and $D_d$ stand for the distances of the source
and the lens from the observer, and $D_{ds}$ represents the lens-source 
distance, as shown in the Fig. 1.  Thus, the lens equation may be expressed as
\be
\tan\beta =  \tan\theta - \alpha ,
\label{LensEqn}
\ee
where
\be
\alpha \equiv 
    \frac{D_{ds}}{D_s}  \lt[\tan\theta + \tan\lt(\hat{\alpha} - \theta\rt)\rt].
\label{Alpha}
\ee
The lens  diagram gives
\be
\sin\theta = \frac{J}{D_d}.
\ee

A gravitational field deflects a light ray and  causes a change in the
cross-section of a bundle of rays. The magnification of an image is defined
as the ratio of the flux of the image to the flux of the unlensed source.
According to  Liouville's theorem the surface brightness is preserved
in gravitational light deflection. Thus, the magnification of an image
turns out to be the ratio of the solid angles of the image and of the  unlensed
source (at the observer). Therefore, for a circularly symmetric
GL, the magnification of an image is given by
\be
\mu = \lt( \frac{\sin{\beta}}{\sin{\theta}} \ \frac{d\beta}{d\theta} \rt)^{-1}.
\label{Mu}
\ee
The sign of the magnification of an image gives the parity of the image. The
singularities in the magnification in the lens plane are known as {\em critical
curves} (CCs) and the corresponding values in the source  plane are known as
{\em caustics}. Critical images are defined as images of $0$-parity.

The tangential and  radial magnifications are expressed by
\be
\mu_t \equiv \lt(\frac{\sin{\beta}}{\sin{\theta}}\rt)^{-1}, ~ ~ ~ 
\mu_r \equiv \lt(\frac{d\beta}{d\theta}\rt)^{-1}
\label{MutMur}
\ee
and singularities in these give {\em tangential critical curves} (TCCs)
and {\em radial critical curves} (RCCs), respectively; the corresponding
values in the source plane are known as  {\em tangential caustic} (TC)
and {\em radial caustics} (RCs), respectively. Obviously, $\beta = 0$
gives the TC and the corresponding values of $\theta$ are the TCCs.
For small values of angles $\beta$, $\theta$ and $\hat{\alpha}$
equations $(\ref{LensEqn})$ and  $(\ref{Mu})$ yield the
approximate lens equation and  magnification, respectively,
which have been widely used in studying lensing in  a weak gravitational field
\cite{Schetal92}.
\section{Schwarzschild spacetime and the deflection angle}
\label{sec:DefAngle}
The Schwarzschild spacetime is expressed by the line element
\be
ds^2=\lt(1-\frac{2M}{r}\rt)dt^2- \lt(1-\frac{2M}{r}\rt)^{-1} dr^2
      -r^2\lt(d\vartheta^2+\sin^2 \vartheta d\phi^2\rt),
\label{SchMetric}
\ee
where $M$ is the Schwarzschild mass. When this solution is maximally extended
it has an event horizon  at the Schwarzschild radius $R_s = 2 M$.
The  deflection angle $\hat{\alpha}$ for a light ray with 
closest distance of approach $r_o$ is (Chapters $8.4$ and $ 8.5$
in \cite{Wei72})
\be
\hat{\alpha}\lt(r_o\rt) = 2 \  {\int_{r_o}}^{\infty}
 \frac{dr}{r \  \sqrt{\lt(\frac{r}{r_o}\rt)^2  \lt(1-\frac{2M}{r_o}\rt)
-\lt(1-\frac{2M}{r}\rt)}   } - \pi 
\label{AlphaHatR0}
\ee
and  the impact parameter $J$ is
\be
J = r_o \lt(1-\frac{2M}{r_o}\rt)^{-\frac{1}{2}} .
\label{ImpParaR0}
\ee

A timelike hypersurface $\{r = r_0\}$ in a spacetime is  defined as a photon 
sphere  if the Einstein bending angle of a  light ray with the closest 
distance  of 
approach  $r_0$ becomes unboundedly large. For the Schwarzschild metric
$r_0 = 3M$ is the photon sphere and thus  the  deflection angle $\hat{\alpha}$
is finite  for $r_0 > 3M$.

The  Einstein deflection angle for large $r_o$ is\cite{Viretal98}
\be
\hat{\alpha}\lt(r_o\rt) =
 \frac{4M}{r_o} + \frac{4M^2}{{r_o}^2}\lt(\frac{15\pi}{16}-1\rt)
+ . . .  . . \ \ \ .
\label{AlphaHatWkField}
\ee
We mentioned the above result  only for completeness as it is not  much
known in the literature.
As we are interested to study GL due to light deflection in a strong 
graviational field  we will use Eq. $(\ref{AlphaHatR0})$ for any 
further calculations. 
Introducing radial distance defined in terms of the Schwarzschild radius,
\be
x = \frac{r}{2M} , ~ ~ ~
x_o = \frac{r_o}{2M} , 
\label{XX0}
\ee
the deflection angle $\hat{\alpha}$ and the impact paprameter $J$
take the form
\be
\hat{\alpha}\lt(x_o\rt) = 2 \  {\int_{x_o}}^{\infty}
 \frac{dx}{x  \  \sqrt{\lt(\frac{x}{x_o}\rt)^2 
 \lt(1-\frac{1}{x_o}\rt)
-\lt(1-\frac{1}{x}\rt)}} - \pi 
\label{AlphaHatX0}
\ee
and
\be
J = 2M  x_o \lt(1-\frac{1}{x_o}\rt)^{-\frac{1}{2}}.
\label{ImpParaX0}
\ee
In the computations in the following section we require the  first derivative
of the deflection angle $\hat{\alpha}$ with respect to $\theta$. This is
given by  (see in \cite{Viretal98})
\be
\frac{d\hat{\alpha}}{d\theta} =  \hat{\alpha}'\lt(x_o\rt)
                               \frac{dx_o}{d\theta} ,
\label{DAlphaByDTheta}
\ee
where
\be
 \frac{dx_o}{d\theta} = 
\frac{
    x_o \lt(1-\frac{1}{x_o}\rt)^{\frac{3}{2}}
\sqrt{1-\lt(\frac{2M}{D_{d}}\rt)^2 {x_o}^2 \lt(1-\frac{1}{x_o}\rt)^{-1}}
 }
{\frac{M}{D_{d}} \lt(2x_o-3\rt)}
\label{DX0ByDTheta}
\ee
and  the first derivative of $\hat{\alpha}$ with respect to $x_o$ is 
\be
\hat{\alpha}'\lt(x_o\rt) = \frac{3-2x_o}{{x_o}^2\lt(1-\frac{1}{x_o}\rt)}
{\int_{x_o}}^{\infty}
 \frac{\lt(4 x - 3\rt) dx}
{\lt(3 - 2 x\rt)^2 \  x \   \sqrt{\lt(\frac{x}{x_o}\rt)^2 
 \lt(1-\frac{1}{x_o}\rt)
-\lt(1-\frac{1}{x}\rt)}} .
\label{DAlphaHatByX0}
\ee
\section{Lensing with the Galactic supermassive ``black hole''}
It is known that the Schwarzschild GL in a weak gravitational field gives
rise to an Einstein ring when the source, lens and observer are aligned,
and a pair of images (primary and secondary) of opposite parities when
the lens components are misaligned. 
However, when  the lens 
is a massive compact  object  a strong  gravitational field is
``available'' for  investigation. A light ray can pass  close to the 
photon sphere and go around the lens once, twice, thrice, or many times
(depending on the impact parameter $J$ but for $J>3\sqrt{3} M$) before 
reaching  the observer. Thus, a massive 
compact lens gives rise, in addition to the primary and secondary images, to a 
large number (indeed, theoretically an infinite sequence) of 
images on both sides of the optic axis. We call these images (which are formed
due to the bending of light through more than $3\pi /2$) {\em relativistic 
images},  as the light rays giving rise to them  pass through a strong 
gravitational field  before reaching the observer.   
We call the  rings  which are 
formed  by bending of light  rays more than $2 \pi$, {\em relativistic 
Einstein rings}.

We model the Galactic supermassive ``black hole'' as a Schwarzschild lens.
This has mass $M  = 2.8 \times 10^6 M_{\odot}$ and the distance $D_d = 
8.5 kpc$\cite{Ricetal98};  therefore, the ratio of the mass to the 
distance  $M/D_d \approx 1.57 \times 10^{-11}$.   We consider a point
source, with  the lens  situated half way between the  source and the
observer, i.e. $D_{ds}/D_s = 1/2$. We allow the angular 
position of the source to change keeping $D_{ds}$  fixed. 

We compute positions and magnifications  of two pairs of outermost relativistic
images as well as the primary and secondary images
for different values of the  angular positions of the  source. These 
are  shown in figures 2 and 3 and Table 1 (for relativistic images)
and in Fig. 4 and Table 2 (for primary and secondary images). The angular
positions of the primary and secondary images as well as the critical curves 
are given in arcseconds;  those for relativistic images as well as relativistic
critical curves are expressed in microarcseconds.

In Fig.2  we show how the positions of outer two relativistic images 
on each side of the optic axis change as the
source position changes. To find the angular positions of images on the
same side of the source we plot $\alpha$ (represented by 
continuous curves on right side of the figure) and 
$\tan\theta-\tan\beta$ (represented by dashed curves)
against $\theta$ for a given
value of the source position $\beta$; the points of intersection give the
image positions (see the right side of the Fig. $3$).
   \begin{figure*}
\hbox{\epsfxsize=8 cm\epsfbox{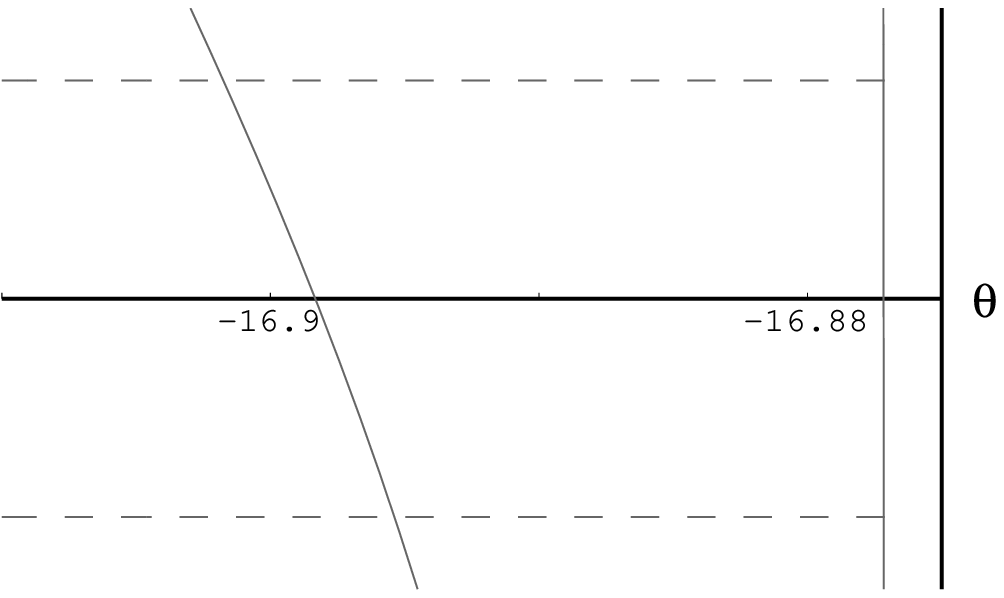}\hskip 0.1 cm
\epsfxsize=9 cm\epsfbox{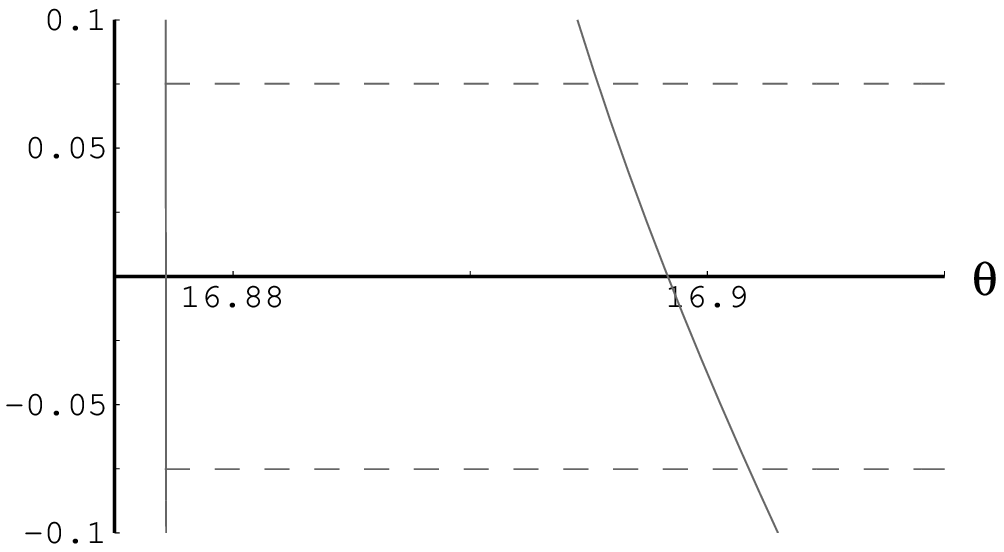}}
 \caption[ ]{ This gives relativistic image positions for a given source
position. $\alpha$ and $\tan\theta-\tan\beta$ are plotted against
the angular position $\theta$ of the image; these are represented by
the continuous and the dashed curves, respectively. For a given position
of the source, the points of intersections of the continuous curves
(the two outermost ones on each side being shown)
with the dashed curves give the angular positions of relativistic images.
The Galactic ``black hole'' (mass $M= 2.8 \times 10^6 M_{\odot}$  
and the distance $D_d =  8.5 kpc$ so that $M/D_d \approx 1.57 
\times 10^{-11}$) serves as the lens, $D_{ds}/D_s = 1/2,$
and $\beta = \mp 0.075$ radian ($\approx \mp 4.29718\degree $).  $\theta$ 
is expressed in microarcseconds. The angular position of a relativistic
image changes very slowly with respect to a change in the source 
position.
}
\label{fig2}
\end{figure*}
\begin{table*}
 \caption[ ]{  Magnifications and positions$^{\rm c}$  of relativistic images }
 \begin{flushleft}
 \begin{tabular}{ccccc}
 \multicolumn{2}{c}{Images on the opposite side of the source }&
         Source position $\beta$        &
 \multicolumn{2}{c}{  Images on the same side of the source} \\
\qquad$\mu^{outer}$&$\mu^{inner}$&
                                   $ $ 
&\qquad$\mu^{inner}$&$\mu^{outer}$ \\
\hline\noalign{\smallskip}
\qquad\phantom{11}$-3.5 \times 10^{-12}$ &\phantom{11} $-6.5 \times 10^{-15}$ 
                      &\qquad $1$ 
          &\qquad $6.5 \times 10^{-15}$ &\phantom{11} $3.5 \times 10^{-12}$ \\
\qquad\phantom{11}$-3.5 \times 10^{-13}$ &\phantom{11} $-6.5 \times 10^{-16}$ 
                      &\qquad $10$ 
          &\qquad $6.5 \times 10^{-16}$ &\phantom{11} $3.5 \times 10^{-13}$ \\
\qquad\phantom{11}$-3.5 \times 10^{-14}$ &\phantom{11} $-6.5 \times 10^{-17}$ 
                      &\qquad $10^2$ 
          &\qquad $6.5 \times 10^{-17}$ &\phantom{11} $3.5 \times 10^{-14}$ \\
\qquad\phantom{11}$-3.5 \times 10^{-15}$ &\phantom{11} $-6.5 \times 10^{-18}$ 
                      &\qquad $10^3$ 
          &\qquad $6.5 \times 10^{-18}$ &\phantom{11} $3.5 \times 10^{-15}$ \\
\qquad\phantom{11}$-3.5 \times 10^{-16}$ &\phantom{11} $-6.5 \times 10^{-19}$ 
                      &\qquad $10^4$ 
          &\qquad $6.5 \times 10^{-19}$ &\phantom{11} $3.5 \times 10^{-16}$ \\
\qquad\phantom{11}$-3.5 \times 10^{-17}$ &\phantom{11} $-6.5 \times 10^{-20}$ 
                      &\qquad $10^5$ 
          &\qquad $6.5 \times 10^{-20}$ &\phantom{11} $3.5 \times 10^{-17}$ \\
 \qquad\phantom{11}$-3.5 \times 10^{-18}$ &\phantom{11} $-6.5 \times 10^{-21}$ 
                      &\qquad $10^6$ 
          &\qquad $6.5 \times 10^{-21}$ &\phantom{11} $3.5 \times 10^{-18}$ \\
 \noalign{\smallskip}
 \noalign{\smallskip}
 \end{tabular}
 \end{flushleft}
\begin{list}{}{}
\item[$^{\rm a}$] 
     The lens is the Galactic ``black hole'' (mass $M= 2.8 \times 10^6 
     M_{\odot} $  and the distance $D_d =  8.5$ kpc  so that $M/D_d
     \approx 1.57 \times  10^{-11}     $). The ratio of the lens-source 
     distance $D_{ds}$ to the observer-source distance $D_s$ is taken to be 
     $1/2$. Angles are given in microarcseconds.
\item[$^{\rm b}$] 
     $\mu$ is the magnification and the  sign on this refers to the parity of 
     the image.
\item[$^{\rm c}$] 
     For the source positions considered here, the angular positions of two
     pairs of outermost relativistic images are $\approx  \pm 16.898$ and 
     $\approx \pm 16.877$ microarcseconds ( $+$ sign refers to images on the
     same side of the source and $-$ sign refers to images on opposite side of
     the source).
\end{list}
 \end{table*}
Similarly, we plot $- \alpha$ and $-\tan\theta-\tan\beta$ vs. $-\theta$ and
points of intersection give the image positions on the opposite side of
the source (see left side of the Fig. $2$). We have taken $\beta = \mp
0.075$ radian ($\approx \mp 4.29718\degree$). In fact there are a sequence 
of theoretically an  infinite number of 
continuous curves which intersect with a given  dashed curve giving rise
to a sequence of an infinite number of images on both sides of the optic
axis. We have plotted only two sets of such curves
(note that the third set of continuous curves comes to be very close to the 
second set and therefore it is not possible to show them in the same figure)
demonstrating appearance of two relativistic images on both sides of the 
optic axis.  For $\beta=0$
the  points of intersection of the continuous curves with the dashed
curve give a sequence of infinite number of relativistic tangential critical
curves (relativistic Einstein rings). As $\beta$ increases any image on
the same side of source moves away from the optic axis,  whereas any image
on the opposite side of the source moves towards the optic axis. The 
displacement of relativistic images with respect to a change in the
source position is very  small (see Fig. $2$). The two sets of
outermost  relativistic images
are formed at about $17$ microarcseconds from the optic axis.

  In Fig. 3  we plot the tangential magnification $\mu_t$ as well as the
total magnification $\mu$ vs. the image position $\theta$ near the two
outermost relativistic tangential critical curves. The singularities in
$\mu_t$ give the angular radii of the two  relativistic Einstein rings.
In  Fig. 4 we plot the same for the primary-secondary images; the singularity
in $\mu_t$ gives the angular position of the Einstein ring.
The magnification for relativistic images falls extremely fast (as compared
with the case of primary and secondary images) as the source position increases
from  perfect alignment.  
The tangential parity  (sign of $\mu_t$) as well as the total parity (sign of 
$\mu$)  are  positive for all images on the same side of the source and
negative for  all  images on the opposite side of the source. The radial 
parity (sign of $\mu_r$) is  positive for all the images in  Schwarzschild 
lensing. 
   \begin{figure*}
\epsfxsize 17cm
   \epsfbox{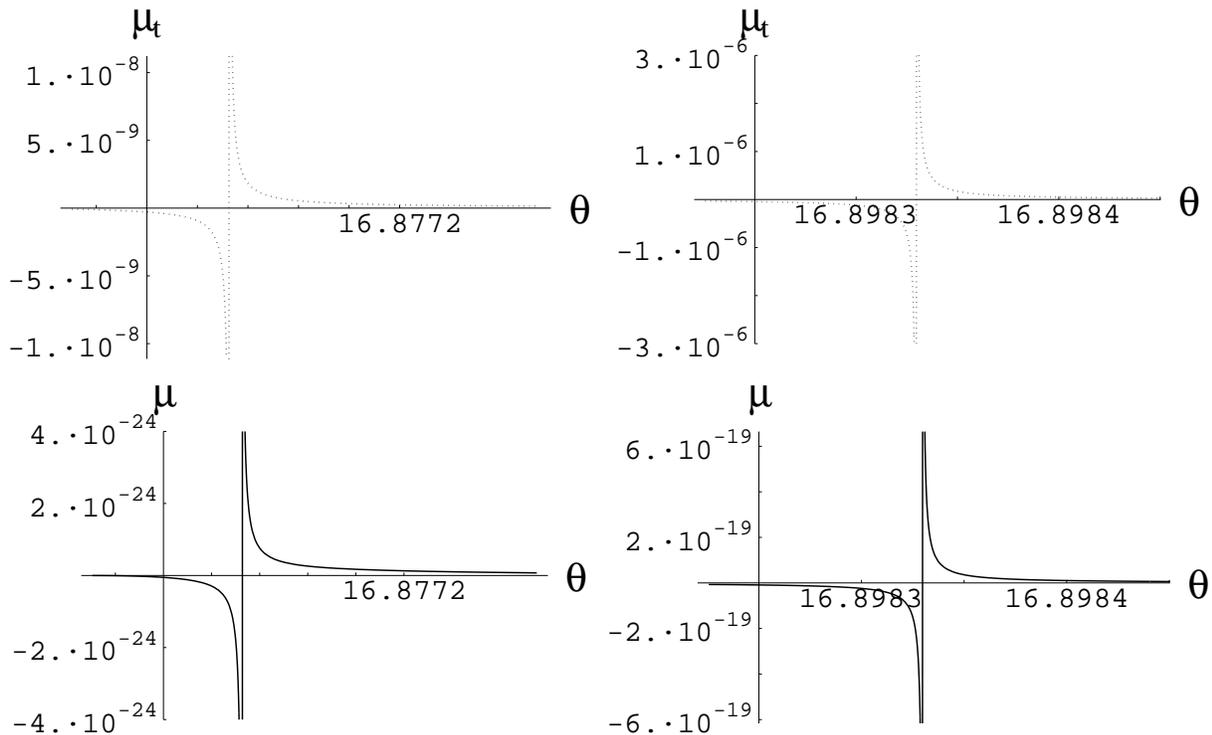}
 \caption[ ]{
The {\em tangential magnification} $\mu_t$ denoted by
dotted curves and the {\em total magnification} $\mu$ denoted by
continuous curves are plotted against the image position
(expressed in microarcseconds) near relativistic tangential critical
 curves. The figures on the right  side give the magnification for the
 outermost relativistic image, whereas those on left side are for
a relativistic image adjacent to the previous one. The lens is the
Galactic ``black hole'' ($M/D_d \approx 1.57 \times 10^{-11}$) and
$D_{ds}/D_s = 1/2$. The singularities in magnifications show
the angular positions of the relativistic tangential critical curves.
The origin of the $\theta$-axis for the figures on the left side is $16.87715$
and for each on right side is $16.89825$.
}
\label{fig3.eps}
\end{figure*}
\begin{table*}
 \caption[ ]{Positions and  magnifications of  primary and secondary images }
 \begin{flushleft}
 \begin{tabular}{crccr}
 \multicolumn{2}{c}{Secondary images }&
          Source position $\beta$     &
 \multicolumn{2}{c}{ Primary images } \\
\qquad$\theta$ & $\mu$ & 
                     $$&
\qquad$\theta$ & $\mu$ \\
 \hline\noalign{\smallskip}
\qquad\phantom{11}$1.157494$ &\phantom{11} $-5787.20 $ 
                &\qquad $10^{-4}$ 
          &\qquad $1.157594$ &\phantom{11} $5788.21 $ \\
\qquad\phantom{11}$1.157045$ &\phantom{11} $-578.27 $ 
                &\qquad $10^{-3}$ 
          &\qquad $1.158045$ &\phantom{11} $579.27 $ \\
\qquad\phantom{11}$1.152555$ &\phantom{11} $-57.38      $ 
                &\qquad $10^{-2}$ 
          &\qquad $1.162555$ &\phantom{11} $58.38 $ \\
\qquad\phantom{11}$1.108619$ &\phantom{11} $-5.30 $ 
                &\qquad $10^{-1}$ 
          &\qquad $1.208624$ &\phantom{11} $6.30 $ \\
\qquad\phantom{11}$0.760918$ &\phantom{11} $-0.23 $ 
                &\qquad $1$ 
          &\qquad $1.760914$ &\phantom{11} $1.23 $ \\
\qquad\phantom{11}$0.529680$ &\phantom{11} $-0.05 $ 
                &\qquad $2$ 
          &\qquad $2.529674$ &\phantom{11} $1.05 $ \\
\qquad\phantom{11}$0.394711$ &\phantom{11} $-0.01 $ 
                &\qquad $3$ 
          &\qquad $3.394704$ &\phantom{11} $1.01$ \\
\qquad\phantom{11}$0.310831$ &\phantom{11} $-0.005 $ 
                &\qquad $4$ 
          &\qquad $4.310823$ &\phantom{11} $1.005$ \\
\qquad\phantom{11}$0.254986$ &\phantom{11} $-0.002 $ 
                &\qquad $5$ 
          &\qquad $5.254977$ &\phantom{11} $1.002$ \\
 \end{tabular}
 \end{flushleft}
\begin{list}{}{}
\item[$^{\rm a}$] 
The same as in Table 1, except  angles are given here in arcseconds.
\end{list}
 \end{table*}
   \begin{figure*}
\epsfxsize 10cm
   \epsfbox{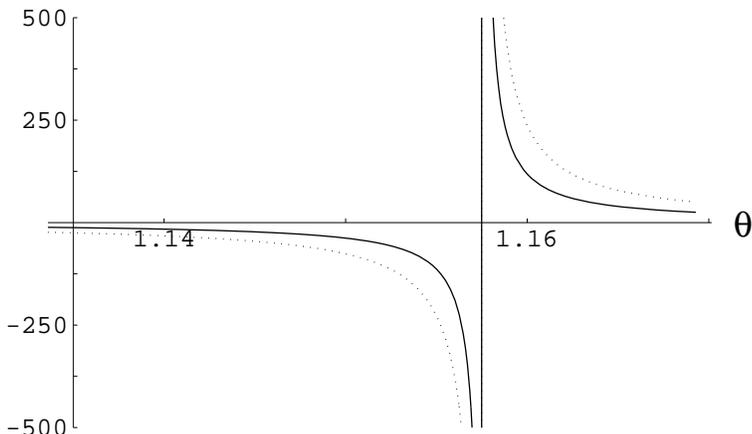}
 \caption[ ]{
The {\em tangential  magnification} $\mu_t$ (represented by dotted curves)
and the {\em total magnification} $\mu$ (represented by continuous  curves)
are plotted, near the outermost tangential critical curve, as a function
of the image position (in arcseconds). The singularity gives the position
of the outermost tangential critical curve (angular radius of the Einstein ring).
The Galactic ``black hole'' is the lens and $D_{ds}/D_s$ is taken to be
$1/2$.
}
\label{fig4.eps}
\end{figure*}
\begin{table*}
\caption[ ]{Einstein and relativistic Einstein rings}
\begin{flushleft}
\begin{tabular}{lllll}
Rings &   $\theta_E$ &  $\hat{\alpha}$ &$\frac{r_o}{2M}$ \\
\noalign{\smallskip}
\noalign{\smallskip}
\hline\noalign{\smallskip}
Einstein ring                 &  $1.157544$ $arcsec$   & $2.315089$ $arcsec$  & $178193$     \\
Relativistic Einstein ring I  &  $16.898$ $\mu$$as$ & $2\pi+33.80$ $\mu$$as$  
                              & $1.545115$\\
Relativistic Einstein ring II &  $16.877$ $\mu$$as$ & $4\pi+33.75$ $\mu$$as$ 
                              & $1.501875$\\
\noalign{\smallskip}
\end{tabular}
\end{flushleft}
\begin{list}{}{}
  \item[$^{\rm a}$]
         The same as ({\rm a}) of Table 1, except $arcsec$ and $\mu as$ 
         used here  refer to arcseconds and microarcseconds, respectively.
         $\theta_E$ stands for  the angular positions of tangential critical 
         curves.
\end{list}
\end{table*}
    In Table 3 we give the angular radii $\theta_E$ of the Einstein and two
relativistic Einstein rings. We also give the corresponding values for the 
deflection angle $\hat{\alpha}$ and the closest distance of approach
$x_o$ for the light rays giving rise to these rings. We define an ``effective
deflection angle'' $\hat{\alpha} - 2 \pi $ times the number of revolution
the light ray has made before reaching the observer. Table 3 shows that the 
effective deflection angle for a ring decreases with the decrease in its 
angular radius, which is expected from the geometry of the lens diagram. The 
same is true for images on the same side of the optic axis, i.e. the effective
deflection angle is less for images closer to the optic axis.
\newpage
The supermassive ``black holes'' at the centres of $NGC3115$ 
and $NGC4486$ have  $M/D_d \approx 1.14 \times 10^{-11}$ and $1.03 
\times 10^{-11}$, resepctively\cite{Ricetal98}, which are  very close to the 
case of the Galactic ``black hole'' we 
have studied. Therefore, if we study lensing with these ``black holes''
keeping  $D_{ds}/D_s = 1/2$, we will get approximately the
same results.  The
angular radius of the Einstein ring in the Schwarzschild lensing 
 is expressed by $\theta_E = \{4M D_{ds}/(D_d D_s)\}^{1/2}$. 
For a source  with $D_{ds} < D_s$ one has $0 < \left( D_{ds}/D_s\right) <1$. 
If we consider  $D_{ds}/D_s$ different than $1/2$
the  magnitude of the Einstein ring can  easily be estimated.
As relativistic images are formed  due to light deflection
close to $r_o = 3M$, their  angular positions will be  very
much less sensitive to a change in the value of $D_{ds}/D_s$.
We have considered the sources for $D_d < D_s$; however, sources with 
$D_d > D_s$ will also be lensed and will also give rise to relativistic 
images.  Thus, all the sources of the universe will be mapped as relativistic 
images in the vicinity of the black hole (albeit as very faint images). 
Gravitational lensing with  stellar-mass black holes
will also give rise to  relativistic images; however, unlike in the
case of  supermassive ``black hole'' lensing,  these images
will not be resolved from their primary and secondary images with present
observational facilities. 

\section{Relativistic images as test for general relativity in strong
         gravitational field}
For the Galactic ``black hole'' lens,  Fig. 2 and Table 1 (see caption
{\rm (c)} )  show the angular positions  of  the two outermost sets of 
relativistic images (two images on each side  of the optic axis) 
when  a source  position is given. 
In fact,  there is a sequence of a large number of relativistic Einstein rings
when the source, lens and observer are perfectly aligned, and when the 
alignment is ``broken'' there is a  sequence of large number of relativistic
images on both sides of the optic axis.  However, for a 
given source position their  magnifications decrease very fast as the 
angular position $\theta$ decreases (see Table 1), and therefore the outermost
set of  images, one on each side of the optic axis, is  observationally
the most  significant.  The angular separations among relativistic images are
too small to be resolved with presently available instruments and therefore
all these images would be at the same position; however, these 
relativistic images  will  be resolved from the primary and secondary images
and thus resolution is not a problem for observation of  relativistic 
images.

    If we observe  a full or  ``broken'' Einstein ring  near 
the centre of a massive dark object at the centre of a galaxy with a faint
relativistic  image of the same source at the centre of the ring, we would 
expect that the  central (relativistic)  image would
disappear after a short period of time. If seen, this would be a great success 
of the  general  theory of relativity in a strong gravitational field. 

Observation  of  relativistic images  would also give  an upper  bound on the
compactness of the lens.  To get a relativistic image a light ray  has to 
suffer  a deflection by an angle $\hat{\alpha} > 3 \pi/2$. For the closest 
distance of approach $r_o = 3.208532 M$ the deflection angle $\hat{\alpha} 
= 269.9999 \degree $ and therefore $r_o/M = 3.208532$ can be considered 
as an upper bound to the compactness of the lens.
The fact that the magnification of a relativistic image decreases very fast
as the source position increases from its perfect alignment with the lens and
observer can be exploited to give a better estimate of the compactness of the
lens. For  the lens system considered in section four, the outermost
relativistic Einstein ring has angular radius about $16.898$ microarcseconds
and this is formed due to light rays bending at the closest distance of 
approach $r_o \approx 3.09023 M_{\odot}$ (see Table $3$). As a relativistic 
image can be observed only  very close to a relativistic TCC, the above value 
of the $r_o/M$ gives an estimate  of compactness of the massive  dark object.

There are some serious difficulties hindering the observation of 
the primary-secondary image pair near a galactic centre; the observation
of relativistic images is even much more difficult. The 
extinction of electromagentic radiation near the line of sight to galactic 
nuclei would be  appreciable; the smaller the wavelength, the larger  the 
extinction. The interstellar scattering and radiation at several frequencies
from the material accreting on the ``black hole'' would make these observations
more difficult. Due to these obstacles no lensing event near a galactic centre
has been observed till now, but it seems this is a very worthwhile project.

There are some additional difficulties for observing relativistic images.
First, these images are very much demagnified unless the source, lens
and observer are highly aligned.
When the source position $\beta$ 
decreases  the magnification increases rapidly and therefore one may  possibly
get observable relativistic images, but only if  the source, lens and observer 
are highly aligned ($\beta << 1$ microarcsecond) and the source has a large
surface brightness. Quasars and supernovae  would be ideal sources for 
observations of  relativistic images. The number of observed quasars is low 
(about $10^4$, see in  \cite{Wam98}) and therefore the probability that a 
quasar will be highly  aligned along the direction of any galactic centre 
of observed galaxies is  extremely  small.  Similarly, there is a very small 
probability that a supernova  will be strongly aligned with any galactic 
centre.  We considered a normal star in the Galaxy to be a point
source (note that we took $D_{ds}/D_s = 1/2$). We cannot
use the point source approximation  when such a source is very close to the
caustic ($\beta = 0$) and therefore studies of extended source  lensing  are 
needed. Second, if relativistic images were observed it  would be for a short 
period of time  because the magnification decreases very fast with  increase 
in the source  position; however, the time scale for observation of 
relativistic images will be greater for lensing of more distant sources.
It is highly improbable that the relativistic images 
would in fact be observed in a short observing period and a long term
project to search for such images would not have reasonable probability
of success. Nevertheless the 
possibility remains that such images might be detected through lucky
observations in the vicinity  of galactic centers.
\section{Summary}
  We obtained  a lens equation which allows an arbitrary large 
values of the deflection angle and used the deflection angle expression
for the Schwarzschild metric obtained by Weinberg\cite{Wei72}. This gives
the bending angle of a light 
ray passing through the  Schwarzschild  gravitational field for a  closest 
distance of approach $r_o$  in the range $3M < r_o < \infty$. Using this we 
studied GL due to the Galactic ``black hole''  in a  strong  
gravitational field. 

Apart from a pair of images (primary and secondary) which are observed due to
light deflection in a weak gravitational field, we  find a sequence of large
number of relativistic images on both sides of the optic axis due to large
deflections of light in a strong gravitational field near the photon sphere
$r_o = 3M$. Among these relativistic images, the outermost pair is 
observationally the most important. Though these relativistic images are
resolved from the primary and secondary images, there are serious difficulties
in observing them. However, if  it were to succeed it would be a great triumph 
of the general  theory  of  relativity and would also provide  valuable  
information about the nature of massive dark  objects. 
Observations  of relativistic images  would confirm the Schwarzschild geometry 
close  to the event horizon; therefore these would strongly support the black 
hole  interpretation of the lensing object.

In the  investigations in this paper we  modelled the massive compact 
objects as Schwarzschild lens. However, it is worth  investigating Kerr  
lensing to see the effect of rotation on lensing in strong gravitational
field, especially  when the lens has large intrinsic angular momentum to 
the mass ratio. There have been some studies of  Kerr weak field lensing 
(see Rauch and  Blandford\cite{RB94} and references therein). In passing,
it is worth mentioning that any spacetime endowed with a photon sphere
(as defined in Section  $\ref{sec:DefAngle}$) and acting as a gravitational 
lens would give rise to relativistic images.

\begin{acknowledgements}
Thanks are due to H. M. Antia,  M. Dominik, J. Kormendy, J. Lehar, and 
D. Narasimha  for helpful correspondence, and J. Menzies and P. Whitelock
for helpful discussions on the visibility of images.  This research was 
supported by FRD, S. Africa.  
\end{acknowledgements}

\end{document}